\documentclass[pra,twocolumn,shownopacs,amssymb]{revtex4}
\usepackage{graphicx}
\usepackage{amsmath}

\begin{document}

\title{Exposing the quantum geometry of spin-orbit coupled Fermi superfluids}

\author{M. Iskin}
\affiliation{Department of Physics, Ko\c{c} University, Rumelifeneri Yolu, 
34450 Sar\i yer, Istanbul, Turkey}

\date{\today}

\begin{abstract}

The coupling between a quantum particle's intrinsic angular momentum 
and its center-of-mass motion gives rise to the so-called helicity states that 
are characterized by the projection of the spin onto the direction of momentum.
In this paper, by unfolding the superfluid-density tensor into its intra-helicity 
and inter-helicity components, we reveal that the latter contribution is directly 
linked with the total quantum metric of the helicity bands. 
We consider both Rashba and Weyl spin-orbit couplings across the BCS-BEC 
crossover, and show that the geometrical inter-helicity contribution is responsible 
for up to a quarter of the total superfluid density. We believe this is one of 
those elusive effects that may be measured within the highly-tunable 
realm of cold Fermi gases.

\end{abstract}


\maketitle

\section{Introduction}
\label{sec:intro}
The earliest association between the physical property of a quantum system 
and the topological structure of its underlying Bloch bands was made back 
in 1980's~\cite{berry89}, soon after the realization that the observed quantization 
of Hall conductivity was nothing but a smoking-gun evidence for its quantum 
geometric origins. In particular, it turns out that the integral of the Berry 
curvature (defined just below) in two-dimensional electron systems is a 
topological invariant of the system, and that its quantized value is proportional 
to the Hall conductivity of a band insulator. Bearing this hindsight in mind, 
not only the topological but also the geometrical structure of the Bloch 
bands have been playing ever-increasing roles in modern quantum physics
~\cite{thouless98, niu10, hasan10, qi11, sinova15, chiu16, bansil16, haldane17}, 
where more and more physical phenomena are proposed to find their 
roots in the so-called geometric quantum mechanics~\cite{brody01}.

In order to specify these topological and geometrical ideas one needs to
introduce the quantum geometric tensor of a given Bloch 
band~\cite{provost80, berry89}, which is also referred to as the Fubini-Study 
metric tensor in the broader context of differential geometry. For this purpose,
let us consider a Bloch Hamiltonian density $H_{0\mathbf{k}}$, 
and label its single-particle energy eigenvalues $\epsilon_{i\mathbf{k}}$ 
and energy eigenstates $| i \mathbf{k} \rangle$ by the band index $i$ and 
momentum $\mathbf{k} = \sum_\nu k_\nu \boldsymbol{\widehat{\nu}}$ 
in the $D$-dimensional Brillouin zone with coordinates $\nu = \{x, y , z\}$ . 
Since the gauge invariant quantum geometric tensor of the $i$th band
$
Q_{i \mathbf{k}}^{\mu \nu} =   
\langle \partial_{k_\mu} i \mathbf{k} | \partial_{k_\nu} i \mathbf{k}  \rangle 
- 
\langle \partial_{k_\mu} i \mathbf{k} | i \mathbf{k} \rangle 
\langle i \mathbf{k} | \partial_{k_\nu} i \mathbf{k} \rangle
$
is a complex one, it is customary to divide it into two 
gauge invariant tensors in the following way
$
Q_{i \mathbf{k}}^{\mu \nu} =  g_{i \mathbf{k}}^{\mu \nu} 
- (\mathrm{i}/2) F_{i \mathbf{k}}^{\mu \nu}
$
~\cite{provost80, berry89}, where $g_{i \mathbf{k}}^{\mu \nu}$ is the so-called 
quantum metric and $F_{i \mathbf{k}}^{\mu \nu}$ the Berry curvature.
Even though the Berry curvature is omnipresent in nature and most of the 
measurable quantum geometric effects have so far been related to 
it~\cite{berry89, thouless98, niu10, hasan10, qi11, sinova15, chiu16, bansil16, haldane17}, 
the importance of the quantum metric is yet to be recognized 
in the light of recent theoretical proposals on a diverse-range of problems in 
condensed-matter physics~\cite{haldane11, mudry13, roy14, roy15, niu14, 
niu15, atac15, kim15, piechon16,torma15, torma16, torma17a, torma17b}. 
Among them, the most recent connection between the quantum metric of 
the non-interacting Bloch bands and the superfluid (SF) weight tensor, 
i.e., in the context of multi-band attractive Hubbard models, stands out as an 
important milestone for our fundamental understandings of superfluidity and 
superconductivity~\cite{torma15, torma16, torma17a, torma17b}. 
For instance, in marked contrast with the single flat-band systems where 
superfluidity is strictly forbidden, it may succeed in a flat-band in the presence 
of other bands (e.g., the Lieb lattice) as a direct result of the geometric effects 
through the inter-band tunnelings~\cite{torma15, torma16}.
In addition, the geometrical contribution to the supercurrent~\cite{torma17a} 
also gives further insight into the baffling controversy around the 
superconductivity of graphene without supercurrent in the vicinity of its 
Dirac points~\cite{kopnin08, kopnin10}.

Motivated by these theoretical proposals~\cite{torma15, torma16, torma17a, torma17b}, 
here we explore the experimental feasibility of its counterpart effect in the SF 
density of spin-orbit coupled Fermi gases~\cite{zhou12, he12a, he12b, gong12}, 
for which the coupling between the intrinsic spin and orbital motion gives rise 
to the so-called helicity states that are characterized by the projection of the spin 
onto the direction of momentum. For this purpose, we first split the SF-density 
tensor into two contributions depending on their physical origin, i.e., while the 
intra-helicity contribution has the conventional form~\cite{denteneer93} 
determined solely by the corresponding helicity spectrum and takes the real
intra-band processes into account, the inter-helicity one accounts for the 
virtual inter-band processes and is directly linked with the total quantum 
metric of the helicity bands. 
We then consider both Rashba and Weyl spin-orbit couplings (SOCs) across 
the BCS-BEC crossover, and show that the geometrical inter-helicity contribution 
is responsible for up to a quarter of the total SF density. Given the recent
realizations of 2D SOCs in atomic Bose and Fermi 
gases~\cite{wu16, sun17, huang16, meng16}, measuring the quantum geometry 
of their helicity bands would endorse the elusive quantum metric to the level of, 
and with arguably as far-reaching impact on modern physics as, 
the Berry curvature.

\section{BCS mean-field theory}
\label{sec:bcs}
Assuming a pseudospin-$1/2$ Fermi SF with an equal number of $\uparrow$ 
and $\downarrow$ components, we start with the BCS mean-field description
of stationary Cooper pairs with zero center-of-mass momentum.  A compact 
way to write this model Hamiltonian (in units of $\hbar = 1$) 
is~\cite{zhou12, he12a, he12b, gong12, iskin11}
\begin{align}
H = \frac{1}{2} \sum_\mathbf{k} &\Psi_\mathbf{k}^\dagger 
\left(
\begin{array}{cc}
  \xi_{\mathbf{k}} \sigma_0 + \mathbf{d}_\mathbf{k} \cdot \boldsymbol{\sigma} & \mathrm{i} \Delta \sigma_y \\
  - \mathrm{i} \Delta \sigma_y & - \xi_{\mathbf{k}} \sigma_0 + \mathbf{d}_\mathbf{k} \cdot \boldsymbol{\sigma}^* \\
\end{array}
\right)
\Psi_\mathbf{k} \nonumber \\
& + \sum_\mathbf{k} \xi_\mathbf{k} + \frac{\Delta^2}{U},
\label{eqn:ham}
\end{align}
where the spinor operators
$
\Psi_\mathbf{k}^\dagger = (\psi_\mathbf{k}^\dagger  \,\,\, \psi_{-\mathbf{k}})
$
with 
$\psi_\mathbf{k}^\dagger = 
(\psi_{\uparrow \mathbf{k}}^\dagger \,\,\, \psi_{\downarrow \mathbf{k}}^\dagger)
$
create $\sigma=\{\uparrow, \downarrow\}$ fermions with $\pm \mathbf{k}$ momentum, 
the shifted dispersion $\xi_\mathbf{k} = \epsilon_\mathbf{k} - \mu$ describes a 
free Fermi gas with the single-particle energy $\epsilon_\mathbf{k} = k^2/(2m)$ 
and the chemical potential $\mu$, and the BCS mean-field 
$
\Delta = U \langle \psi_{\uparrow \mathbf{k}} \psi_{\downarrow -\mathbf{k}} \rangle
$
is taken as a real parameter without the loss of generality. Here, $U \ge 0$ is the strength 
of the contact interaction and $\langle \cdots \rangle$ denotes the thermal average. 
Furthermore, $\sigma_0$ is the $2 \times 2$ identity matrix, and 
$
\boldsymbol{\sigma} = \sum_\nu \sigma_\nu \boldsymbol{\widehat{\nu}}
$
is a vector of Pauli spin matrices in such a way that
$
\mathbf{d}_\mathbf{k} = \sum_\nu \alpha_\nu k_\nu \boldsymbol{\widehat{\nu}} 
$
corresponds to a Weyl SOC when $\alpha_\nu = \alpha$ for all $\nu = \{x,y,z\}$,
and to a Rashba SOC when $\alpha_z = 0$. Here, $\boldsymbol{\widehat{\nu}}$ 
is a unit vector along the $\mathbf{\nu}$ direction, and we choose $\alpha \ge 0$ 
without losing generality.

Since this Hamiltonian and its numerous variations have been well-studied 
in the recent cold-atom literature, we simply quote the self-consistency 
equations for $\Delta$ and $\mu$~\cite{zhou12, he12a, he12b, gong12, iskin11},
\begin{align}
\label{eqn:op}
\frac{1}{U} &= \frac{1}{2\mathcal{V}_D} \sum_{s \mathbf{k}} \frac{\mathcal{X}_{s \mathbf{k}}}{2E_{s \mathbf{k}}}, \\
n &= \frac{1}{2\mathcal{V}_D} \sum_{s \mathbf{k}} \left(
1 - \frac{\xi_{s\mathbf{k}}}{E_{s\mathbf{k}}} \mathcal{X}_{s\mathbf{k}} 
\right),
\label{eqn:num}
\end{align}
where $\mathcal{V}_D$ corresponds to the area $A$ in 2D and volume $V$ in 3D, 
$s = \pm$ labels the helicity bands,
$
\mathcal{X}_{s\mathbf{k}} = \tanh [E_{s\mathbf{k}}/(2T)]
$
is a thermal factor with the Boltzmann constant $k_B$ set to unity 
and $T$ the temperature,
$
\xi_{s\mathbf{k}} = \epsilon_{s\mathbf{k}} - \mu
$
is the shifted dispersion for the $s$-helicity band with
$
\epsilon_{s\mathbf{k}} = \epsilon_\mathbf{k} + sd_\mathbf{k}
$
and the strength of the SOC
$
d_\mathbf{k} = |\mathbf{d}_\mathbf{k}|,
$
and
$
E_{s\mathbf{k}} = (\xi_{s\mathbf{k}}^2 + \Delta^2)^{1/2}
$
is the energy spectrum of the quasiparticles for the corresponding helicity band.
Here, the number equation~(\ref{eqn:num}) for the density $n = N/\mathcal{V}_D$ 
of particles follows from
$
N = \sum_\mathbf{k} \langle \psi_\mathbf{k}^\dagger \psi_\mathbf{k} \rangle.
$
Given that the model Hamiltonian is effectively a two-band one with a single 
$\mathbf{k}$-independent order parameter, and the time-reversal symmetry 
is also manifest, we argue in this paper that spin-orbit coupled Fermi SFs 
may promise one of the ideal test-beds for the exploration of the recently proposed 
quantum geometric effects~\cite{torma15, torma16, torma17a, torma17b}.

\section{Superfluid-density tensor}
\label{sec:sfd}
As a counterpart to the geometric effects in the SF-weight tensor of multi-band 
attractive Hubbard models~\cite{torma15, torma16, torma17a, torma17b}, 
here we study SF-density tensor $\rho_{\mu\nu}$ of a continuum model in 
the context of spin-orbit coupled Fermi gases. 
Following Ref.~\cite{torma17a, torma17b}, we unravel the intra-helicity 
and inter-helicity contributions to
$
\rho_{\mu\nu} = \rho_{\mu\nu}^{intra} + \rho_{\mu\nu}^{inter}
$
as follows
\begin{align}
\label{eqn:intra}
\rho_{\mu\nu}^{intra} &= \frac{m\Delta^2}{2\mathcal{V}_D} \sum_{s \mathbf{k}}
\left(
\frac{\mathcal{X}_{s\mathbf{k}}}{E_{s\mathbf{k}}^3} 
- \frac{\mathcal{Y}_{s\mathbf{k}}}{2T E_{s\mathbf{k}}^2}
\right)
\frac{\partial \xi_{s\mathbf{k}}} {\partial k_\mu}
\frac{\partial \xi_{s\mathbf{k}}} {\partial k_\nu},
\\
\rho_{\mu\nu}^{inter} &= - \frac{m \Delta^2}{\mathcal{V}_D} \sum_{s \mathbf{k}}
\frac{d_\mathbf{k} \mathcal{X}_{s\mathbf{k}}}{s \xi_\mathbf{k} E_{s\mathbf{k}}} 
g_\mathbf{k}^{\mu\nu},
\label{eqn:inter}
\end{align}
where 
$
\mathcal{Y}_{s\mathbf{k}} = \mathrm{sech}^2 [E_{s\mathbf{k}}/(2T)]
$
is a thermal factor, and
$
g_\mathbf{k}^{\mu\nu} = \sum_s g_{s \mathbf{k}}^{\mu\nu}
$
is the total quantum metric of the helicity bands. Here, while 
$\rho_{\mu \nu}^{intra}$ is finite unless $\Delta = 0$, $\rho_{\mu \nu}^{inter}$ 
is finite unless $\alpha = 0$ together with $\Delta = 0$.

We recall that the quantum metric of a given Bloch band is generally defined 
by the energy spectrum $\epsilon_{i\mathbf{k}}$ of the Hamiltonian and its 
corresponding eigenfunctions $|i \mathbf{k} \rangle$ in a highly non-trivial 
way~\cite{provost80, berry89}. This can be illustrated by combining the generic 
definition of the metric
$
g_{i\mathbf{k}}^{\mu\nu} = \mathrm{Re}
[\langle \partial_{k_\mu} i \mathbf{k} | 
(\mathbb{I} - | i \mathbf{k} \rangle \langle i \mathbf{k} |) 
| \partial_{k_\nu} i \mathbf{k} \rangle]
$
for the $i$th band together with the completeness relation
$
\mathbb{I} = \sum_i |i \mathbf{k} \rangle \langle i \mathbf{k}|
$
for a given $\mathbf{k}$ state, 
leading to an equivalent but numerically much more practical expression
$
g_{i \mathbf{k}}^{\mu\nu} = \mathrm{Re} 
\sum_{j \lbrace \ne i \rbrace} 
\langle i \mathbf{k}| 
\partial_{k_\mu} H_{0\mathbf{k}}
| j \mathbf{k} \rangle
\langle j \mathbf{k}| 
\partial_{k_\nu} H_{0\mathbf{k}}
| i \mathbf{k} \rangle
/ (\epsilon_{i \mathbf{k}} - \epsilon_{j \mathbf{k}})^2.
$
In particular application to our model, the single-particle problem is determined by
the wave equation
$
H_{0\mathbf{k}} |s \mathbf{k}\rangle = \epsilon_{s \mathbf{k}} |s \mathbf{k}\rangle,
$
where
$
H_{0\mathbf{k}} = \epsilon_\mathbf{k} \sigma_0 + \mathbf{d}_\mathbf{k} \cdot \boldsymbol{\sigma}
$
is the Hamiltonian density, giving rise to two helicity bands indexed by 
$s = \pm$ as long as $\alpha \ne 0$. Thus, we find
$
g_{+,\mathbf{k}}^{\mu \nu} = g_{-, \mathbf{k}}^{\mu\nu}
$
so that
$
g_\mathbf{k}^{\mu\nu} = 
\alpha_\mu\alpha_\nu (d_\mathbf{k}^2 \delta_{\mu\nu}
- \alpha_\mu \alpha_\nu k_\mu k_\nu)/(2d_\mathbf{k}^4)
$
with $\delta_{ij}$ the Kronecker delta, and also that the total quantum 
metric can also be represented as
$
g_\mathbf{k}^{\mu\nu} = 
\partial_{k_\mu} \widehat{\mathbf{d}}_\mathbf{k} 
\cdot \partial_{k_\nu} \widehat{\mathbf{d}}_\mathbf{k}/2,
$
where
$
\widehat{\mathbf{d}}_\mathbf{k} = \mathbf{d}_\mathbf{k}/d_\mathbf{k}
$
is a unit vector along the SOC field. In comparison, we find 
$
F_{s \mathbf{k}}^{\mu \nu} = sF_{\mathbf{k}}^{\mu\nu}
$
with
$
F_{\mathbf{k}}^{\mu\nu} = 
(\partial_{k_\mu} \widehat{\mathbf{d}}_\mathbf{k} 
\times \partial_{k_\nu} \widehat{\mathbf{d}}_\mathbf{k})
\cdot \widehat{\mathbf{d}}_\mathbf{k}/2
$
for the corresponding Berry curvatures, where each one of its components 
is determined by the quantum metric,
$
|F_{\mathbf{k}}^{\mu\nu}|
= (g_{\mathbf{k}}^{\mu\mu} g_{\mathbf{k}}^{\nu\nu}
- g_{\mathbf{k}}^{\mu\nu} g_{\mathbf{k}}^{\nu\mu})^{1/2},
$
up to a $\mathbf{k}$-dependent sign.

Prior to studying the interplay between the intra-helicity and
inter-helicity contributions, let us briefly sketch how their tangled 
sum reproduces the familiar expressions reported in the recent cold-atom 
literature~\cite{zhou12, he12a, he12b, gong12}. 
For this purpose, we first recast the conventional number 
Eq.~(\ref{eqn:num}) via an integration by parts, i.e.,
$
N = - (1/2) \sum_{s\mathbf{k}} k_\nu \partial_{k_\nu} 
(1 -  \xi_{s\mathbf{k}} \mathcal{X}_{s\mathbf{k}}/E_{s\mathbf{k}}),
$
into an equivalent but somewhat unfamiliar form
$
N = (1/2) \sum_{s\mathbf{k}} k_\nu (\partial \xi_{s\mathbf{k}}/\partial {k_\nu}) 
[\Delta^2 \mathcal{X}_{s\mathbf{k}}/E_{s\mathbf{k}}^3
+ \xi_{s\mathbf{k}}^2 \mathcal{Y}_{s\mathbf{k}}/(2T E_{s\mathbf{k}}^2)].
$
This alternative expression holds for any $\nu$ as long as $\Delta \ne 0$, 
and we attested its accuracy in our numerics as well. Plugging it into 
Eq.~(\ref{eqn:intra}) and taking the $\alpha \to 0$ limit, we find that,
$
\rho_{\mu\nu}^{intra} = n \delta_{\mu\nu} 
- \sum_\mathbf{k} k_\mu k_\nu \mathcal{Y}_{\mathbf{k}}/(2mAT)
$ 
reduces to the conventional expression for a continuum Fermi 
SF~\cite{denteneer93}, and that Eq.~(\ref{eqn:inter}) vanishes as the 
helicity bands unite in this limit.
This leads to $\rho_{\mu\nu} = n \delta_{\mu\nu}$ at $T = 0$, i.e.,
the entire Fermi gas is a SF in the ground state as soon as $\Delta \ne 0$.
When $\alpha \ne 0$, in order to attain the precise form of the SF density 
of, e.g., a 2D Fermi gas with Rashba SOC~\cite{he12a},
$
\rho_0 = n - (m/A)\sum_{s\mathbf{k}} 
[\alpha(\Delta^2 + \xi_\mathbf{k} \xi_{s\mathbf{k}}) \mathcal{X}_{s\mathbf{k}} 
/(4 s k \xi_\mathbf{k} E_{s\mathbf{k}})
+ (k/m + s\alpha)^2 \mathcal{Y}_{s\mathbf{k}}/(8T)],
$
where $k = (k_x^2 + k_y^2)^{1/2}$, we perform yet another integration by 
parts on the following term,
$
\Delta^2 \sum_{\mathbf{k}} (\alpha^2 + s\alpha k/m) 
\mathcal{X}_{s\mathbf{k}}/E_{s\mathbf{k}}^3 = s\alpha A k_c
- s\alpha\sum_\mathbf{k} \xi_{s\mathbf{k}} \mathcal{X}_{s\mathbf{k}}/(k E_{s\mathbf{k}})
- \sum_\mathbf{k} (\alpha^2 + sk\alpha/m) \xi_{s\mathbf{k}}^2 \mathcal{Y}_{s\mathbf{k}}/(2TE_{s\mathbf{k}}^2),
$
in the alternative number equation. Here, the ultraviolet $\mathbf{k}$-space 
cutoff $k_c$ cancels out once summed over $s$. Finally, observing that
$
g_\mathbf{k}^{\mu\nu} = (k^2 \delta_{\mu\nu}-k_\mu k_\nu)/(2k^4)
$
may effectively be replaced with
$
g_\mathbf{k}^{\mu\nu} = \delta_{\mu\nu}/(4k^2),
$
thanks to the even sums over $k_x$ and $k_y$, we eventually arrive at an 
isotropic tensor $\rho_{\mu\nu} = \rho_0 \delta_{\mu\nu}$ with the quoted SF 
density as the prefactor. Similar procedures apply to 3D Fermi SFs with 
Weyl~\cite{he12b} and Rashba~\cite{zhou12} SOCs. In addition, we also 
verified that all of our numerical results for $\rho_{\mu\nu}$ benchmark 
perfectly well with the existing literature~\cite{zhou12, he12a, he12b, gong12}.

To the best of our knowledge, the significance of the inter-helicity contribution
Eq.~(\ref{eqn:inter}) to the SF-density tensor has entirely gone unnoticed in the 
physics literature. Having firmly established its geometric origin~\cite{footnote}, 
next we explore its relative weight in the whole parameter space starting with 
a 2D Fermi SF with Rashba SOC.

\begin{figure}[htbp]
\includegraphics[scale=0.45]{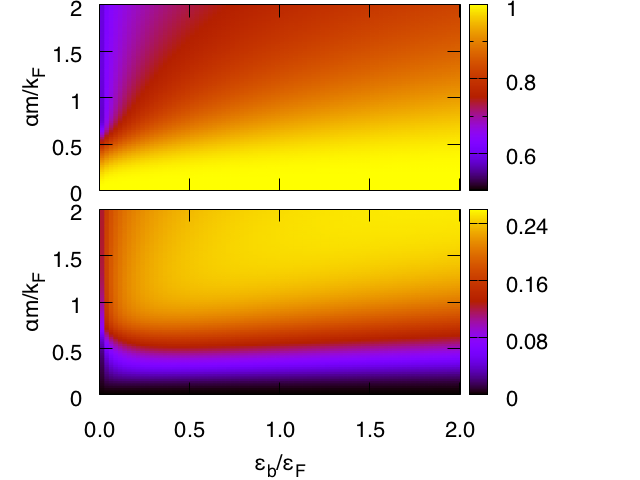}
\caption{(color online)
\label{fig:2DRashba}
A 2D Fermi gas with Rashba SOC is mapped at $T = 0$ in the plane of 
two-body binding energy $\epsilon_b$ and SOC strength $\alpha$. 
The total SF fraction $\rho_0/n$ is shown at the top along with the overall 
inter-helicity fraction $\rho_0^{inter}/\rho_0$ at the bottom.
}
\end{figure}
\subsection{2D Fermi gas with Rashba SOC}
\label{sec:2D}
In line with the cold-atom literature, here we substitute $U$ with the 
two-body binding energy $\epsilon_b \ge 0$ in vacuum via the 
usual relation
$
A/U = \sum_\mathbf{k} 1/(2\epsilon_\mathbf{k} + \epsilon_b).
$
In addition, we set
$
n = k_F^2/(2\pi),
$
and define an effective Fermi momentum $k_F$ and Fermi energy 
$\epsilon_F = k_F^2/(2m)$ as the relevant length and energy scales 
in our numerical calculations. This is in such a way that increasing $U$ 
from $0$ increases $\epsilon_b$ continuously from $0$ to $\infty$.
For instance, a colored map of the ground-state ($T = 0$) SF density
$
\rho_0 = \rho_0^{intra} + \rho_0^{inter}
$ 
is shown in Fig.~\ref{fig:2DRashba} as functions of $\epsilon_b$ and 
$\alpha$, along with the overall inter-helicity fraction $\rho_0^{inter}/\rho_0$. 
See the Appendix for analogous results near the critical BKT transition 
temperature $T = T_{BKT}$, showing that the thermal effects are quite 
negligible for most of the parameter regimes of interest.

In the absence of a SOC when $\alpha = 0$, Fig.~\ref{fig:2DRashba} 
reveals that $\rho_0 = n$ and $\rho_0^{inter} = 0$ for any 
$\epsilon_b > 0$, which is a well-known result in the condensed-matter 
literature~\cite{denteneer93}. On the other hand, increasing $\alpha$ 
from $0$ gradually depletes the SF fraction down to a saturation value 
that is eventually determined by the effective mass of the Cooper 
molecules in the strong-coupling limit, i.e.,
$
\rho_0/n \to 2m/m_B
$
for a weakly-interacting molecular Bose SF.
For instance, it is already known that $m_B/m \to \{2, 4\}$ when 
$m \alpha/k_F \to \{0, \infty\}$~\cite{he12a}, and that the SF 
fraction exhibits a dip value of $1/2$. 
This is barely seen in Fig.~\ref{fig:2DRashba} in a tiny region
when $\epsilon_b/\epsilon_F \to 0$ for sufficiently large $m \alpha/k_F$. 
More interestingly, increasing $\alpha$ from $0$ builds up the overall 
inter-helicity contribution $\rho_0^{inter}/\rho_0$, growing up slowly to 
a maximal value of $0.26$ for the parameters shown. Furthermore, even 
though it is not visible here in a 2D system, as a direct outcome of the 
competition between the intra-helicity and inter-helicity contributions, 
$\rho_0$ evolves non-monotonously with $\alpha$ especially when 
$\epsilon_b/\epsilon_F \ll 1$. Such an interplay is best illustrated in a 
3D Fermi gas as we discuss next.

\begin{figure}[htbp]
\includegraphics[scale=0.45]{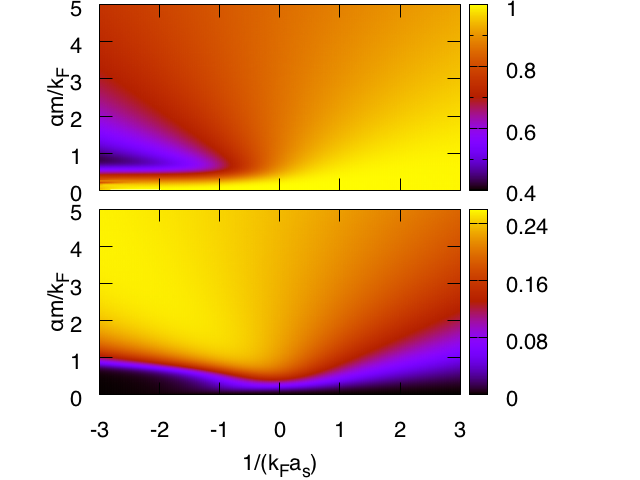}
\caption{(color online)
\label{fig:3DWeyl}
A 3D Fermi gas with Weyl SOC is mapped at $T = 0$ in the plane of 
two-body scattering length $a_s$ and SOC strength $\alpha$. 
The total SF fraction $\rho_0/n$ is shown at the top along with the overall 
inter-helicity fraction $\rho_0^{inter}/\rho_0$ at the bottom.
}
\end{figure}
\subsection{3D Fermi gas with Weyl or Rashba SOC}
\label{sec:3D}
In line with the cold-atom literature, here we substitute $U$ with the two-body 
scattering length $a_s$ in vacuum via the usual relation
$
V/U = -mV/(4\pi a_s) + \sum_\mathbf{k} 1/(2\epsilon_\mathbf{k}),
$
and choose
$
n = k_F^3/(3\pi^2)
$
to define the relevant length and energy scales for the numerical calculations.
This is in such a way that increasing $U$ from $0$ changes the dimensionless 
parameter $1/(k_F a_s)$ continuously from $-\infty$ to $0$ to $+\infty$, for 
which $|a_s| \to \infty$ is commonly referred to as the unitarity.
For instance, a colored map of the ground-state SF density
$
\rho_0 = \rho_0^{intra} + \rho_0^{inter}
$ 
is shown for the Weyl SOC in Fig.~\ref{fig:3DWeyl} as functions of $1/a_s$ 
and $\alpha$, along with the overall inter-helicity fraction $\rho_0^{inter}/\rho_0$.

First of all, it is again already known that $m_B/m \to \{2, 2.32, 6\}$ when 
$
1/(m \alpha a_s) \to \{ +\infty, 0, -\infty \}
$ 
~\cite{iskin11, he12b}, and therefore, we expect the depletion of the SF 
fraction $\rho_0/n$ to saturate around $0.34$ when $1/(k_Fa_s) \ll 0$ and 
around $0.86$ at unitarity for sufficiently large $m \alpha/k_F$.
Our numerical results shown in Fig.~\ref{fig:3DWeyl} nicely recover 
these limits. In addition, similar to a 2D Fermi SF, we find that the overall 
inter-helicity contribution $\rho_0^{inter}/\rho_0$ builds again up to a 
maximal value of $0.26$ for the parameters shown. Furthermore, the rapid 
growth of $\rho_0^{inter}$ on the BCS side of the unitarity leads to a 
non-monotonous evolution of $\rho_0$ with $\alpha$, which is clearly 
visible in a broad region when $1/(k_Fa_s) \lesssim 0$.

\begin{figure}[htbp]
\includegraphics[scale=0.45]{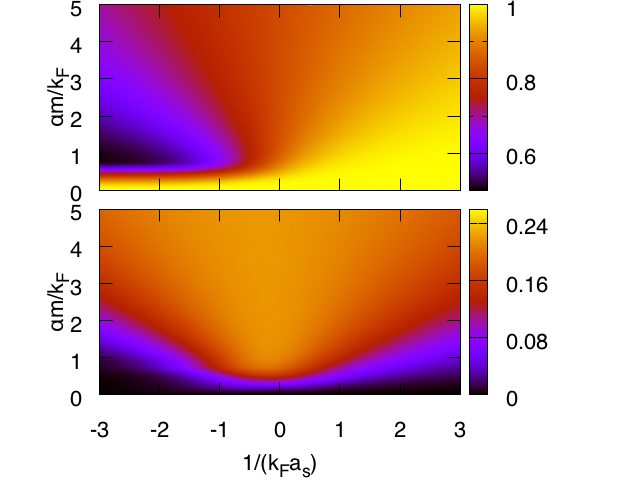}
\caption{(color online)
\label{fig:3DRashba}
A 3D Fermi gas with Rashba SOC is mapped at $T = 0$ in the plane of 
two-body scattering length $a_s$ and SOC strength $\alpha$. 
The total SF fraction $\rho_\perp/n$ is shown at the top along with the overall 
inter-helicity fraction $\rho_\perp^{inter}/\rho_\perp$ at the bottom.
}
\end{figure}

For completeness, we also present a colored map of the in-plane ground-state 
SF density
$
\rho_\perp = \rho_\perp^{intra} + \rho_\perp^{inter}
$ 
for the Rashba SOC in Fig.~\ref{fig:3DRashba}, along with the overall 
inter-helicity fraction $\rho_\perp^{inter}/\rho_\perp$. Note that 
$\rho_{zz} = n$ and $\rho_{zz}^{inter} = 0$ for the entire parameter space 
at $T = 0$, and are not shown. Our numerical results are again in perfect 
agreement with the expected results, for which we already know that
$m_B/m \to \{2, 2.40, 4\}$ when 
$
1/(m \alpha a_s) \to \{ +\infty, 0, -\infty \}
$ 
~\cite{iskin11, zhou12}, leading $\rho_0/n$ to saturate around 
$0.5$ when $1/(k_Fa_s) \ll 1$ and around $0.84$ at unitarity 
for sufficiently large $m \alpha/k_F$. 
In addition, the overall inter-helicity contribution $\rho_0^{inter}/\rho_0$ 
builds up to a maximal value of $0.22$ for the parameters shown. Thus,
in comparison to the Weyl SOC shown in Fig.~\ref{fig:3DWeyl}, 
$\rho_0^{inter}$ is slightly weaker here on the BCS side of the unitarity, 
even though it is rather comparable on the BEC side. 
As the 2D SOCs have recently been created with atomic 
Bose and Fermi gases~\cite{wu16, sun17, huang16, meng16}, our 
predictions in this paper may already be verified in similar setups.
Noting that the quantum metric effects have so far proved to be quite rare and 
elusive in condensed-matter physics~\cite{haldane11, mudry13, roy14, roy15, 
niu14, niu15, atac15, kim15, piechon16,torma15, torma16, torma17a, torma17b}, 
in contrary to the Berry curvature ones that are ubiquitously found in 
nature~\cite{thouless98, niu10, hasan10, qi11, sinova15, chiu16, bansil16, haldane17},
there is no doubt that its cold-atom realization will be one of the landmark 
breakthroughs in modern quantum physics.

\section{Conclusions}
\label{sec:conc}
In summary, while having primarily focused on the spin-orbit coupled atomic 
Fermi SFs but not necessarily limited to it, i.e., in the much broader context 
of superfluidity and superconductivity, here we examined the prospects for 
revealing the quantum geometry of the non-interacting helicity bands 
through measuring the SF density of the system. For this purpose, we first 
divide the SF-density tensor into two contributions based on their physical 
origin, i.e., while the intra-helicity contribution has the conventional form 
determined by the helicity spectrum, the inter-helicity 
one has a geometric origin related to the total quantum metric of the helicity 
bands. We then considered both Rashba and Weyl SOCs across the 
BCS-BEC crossover, and showed that the geometrical contribution accounts 
for up to a quarter of the total SF density. 

Thus, by studying the competition between the intra-band and inter-band 
contributions to the SF density, as well as the hidden role played by the quantum 
metric, our extensive numerical calculations on spin-orbit coupled Fermi gases 
exposed the missing link between the non-monotonous evolution of the SF density 
and the quantum geometry of the helicity bands. This is our main finding in 
this paper. In addition, our work also shed light on the underlying physical 
mechanism behind other non-monotonous effects as the SF density is directly 
related to the mass of the SF carriers. For instance, in the follow-up 
studies~\cite{qm-mass, qm-X, qm-sp}, we have recently showed that the quantum 
metric governs not only the SF density but also many other observables 
including the sound velocity, spin-susceptibility, etc., through renormalizing the 
effective mass of the two-body bound states and Cooper pairs in general. 

Incentivized by the recent creations of 2D SOCs along with the ongoing push 
toward simulating diverse aspects of spin-orbit physics in the cold-atom community~\cite{wu16, sun17, huang16, meng16}, 
we believe realization of such a geometric effect will be one of the long-standing
milestones in modern quantum many-body physics, where not only the 
topology but also the geometry of the underlying band structure play 
ever-increasing roles. As a possible probe, we expect non-monotonic evolutions 
for those SF (normal-state) properties that are inversely proportional to the 
effective mass of the SF carriers (pre-formed pairs).

\begin{acknowledgments}
The author acknowledges support from T{\"U}B{\.I}TAK and the BAGEP award 
of the Turkish Science Academy.
\end{acknowledgments}

\appendix

\section{BKT transition temperature}

\begin{figure}[htbp]
\includegraphics[scale=0.45]{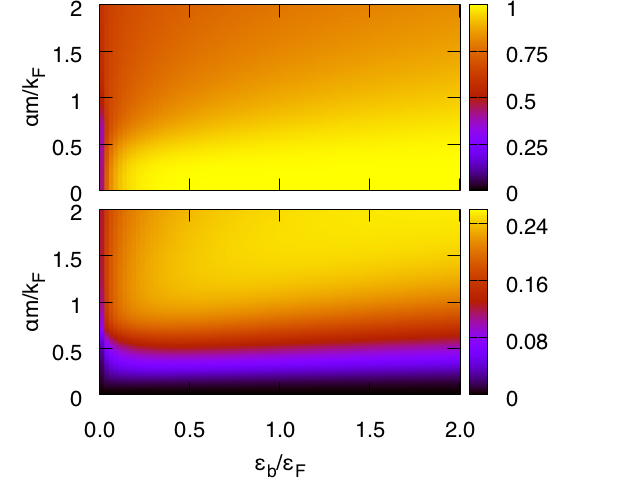}
\caption{(color online)
\label{fig:2DRashbaBKT}
A 2D Fermi gas with Rashba SOC is mapped at $T = T_{BKT}$ in the plane of 
two-body binding energy $\epsilon_b$ and SOC strength $\alpha$. 
The total SF fraction $\rho_0/n$ is shown at the top along with the overall 
inter-helicity fraction $\rho_0^{inter}/\rho_0$ at the bottom.
}
\end{figure}

The BKT transition temperature is determined by the universal BKT relation,
$
T_{BKT} = \frac{\pi}{8m} \sqrt{\rho_{xx} \rho_{yy} - \rho_{xy} \rho_{yx}}
= \pi \rho_0 / (8m),
$
self-consistently with the mean-field order parameter $\Delta$ and the 
corresponding chemical potential $\mu$. As shown in Fig.~\ref{fig:2DRashbaBKT}, 
the geometric effect remains intact even at $T = T_{BKT}$, and it is very much 
similar to that of the ground state that is shown in Fig.~\ref{fig:2DRashba}.

\begin{thebibliography}{99}
\bibitem{footnote}
Note that the local quantum geometry is non-Abelian as the helicity bands 
are degenerate for some $\mathbf{k}$ states in the parameter space, 
e.g., around the origin in our continuum model. However, thanks to the 
$\mathbf{k}$ sums, it turns out that the inter-helicity contribution to some 
of the global properties such as the SF density can be expressed in terms 
of the total quantum metric of the helicity bands as discussed in the text.
\end{thebibliography}
\end{document}